\documentclass[11pt]{article}
\usepackage{amssymb,amsthm,fullpage}
\usepackage{makeidx}
\usepackage{latexsym}
\usepackage{epsf}
\usepackage{epsfig}

\newtheorem{thm}{Theorem}             
\newtheorem{prop}{Proposition}        
\newtheorem{lem}{Lemma}               
\newtheorem{exmp}{Example}            

\def\CSS{\mbox{\sc{Closest Substring}\/}}
\def\CLIQUE{\mbox{\sc{Clique}\/}}

\def\pos{\mbox{\it{pos}\/}}

\def\nat{{\bf N\/}}

\def\P{\mbox{\sc{P}\/}}
\def\NP{\mbox{\sc{NP}\/}}
\def\FPT{\mbox{\sc{FPT}\/}}
\def\W{\mbox{\sc{W}\/}}

\begin{document}
\title{Parameterized Intractability\\ of Motif Search Problems\footnote{An extended abstract of this paper was presented at the 19th International Symposium on Theoretical Aspects of Computer Science (STACS 2002), Springer-Verlag, LNCS 2285, pages 262--273, held in Juan-Les-Pins, France, March~14--16, 2002.}}

\author{
Michael~R.~Fellows\thanks{%
Department of Computer Science and Software Engineering,
University of Newcastle, University Drive, Callaghan 2308, Australia.
Email: mfellows@cs.newcastle.edu.au.
}
\and
Jens Gramm\thanks{%
Wilhelm-Schickard-Institut f\"ur Informatik,
Universit\"at T\"ubingen, Sand 13, D-72076~T\"ubingen,
Fed.\ Rep.\ of Germany.
Email: gramm@informatik.uni-tuebingen.de.
Work was supported by the Deutsche Forschungsgemeinschaft (DFG),
research project ``OPAL'' (optimal solutions for hard problems in
computational biology), NI 369/2-1.}
\and
Rolf Niedermeier\thanks{%
Wilhelm-Schickard-Institut$\,$f\"ur$\,$Informatik,$\,$%
Universit\"at$\,$T\"ubingen,$\,$Sand$\,$13,$\,$D-72076$\,\,$T\"ubingen,$\,$%
Fed.$\,$Rep.$\,$of Germa\-ny.
Email: niedermr@informatik.uni-tuebingen.de.
}}

\date{}
\maketitle 

\begin{abstract}
\noindent
We show that {\sc Closest Substring\/},
one of the most important problems in the field 
of biological sequence analysis, is $\W[1]$-hard when parameterized by
the number~$k$ of input strings (and remains so, even over a binary alphabet).
This problem is therefore unlikely to be solvable in time
$O(f(k)\cdot n^{c})$ for any function~$f$ of~$k$ and constant $c$
independent of $k$.  The problem can therefore be
expected to be intractable, in any practical sense, for $k \geq 3$.
Our result supports the 
intuition that {\sc Closest Substring\/} is computationally much harder than the 
special case of {\sc Closest String\/}, although both problems are 
$\NP$-complete.
We also prove $\W[1]$-hardness for other 
parameterizations in the case of unbounded alphabet size.
Our $\W[1]$-hardness result for {\sc Closest Substring\/} generalizes 
to {\sc Consensus Patterns\/}, a problem of similar 
significance in computational biology.
\end{abstract}

\section{Introduction}
Motif search problems are of central importance for sequence analysis 
in computational molecular biology.
These problems have applications in fields such as genetic drug target 
identification or signal finding (see
\cite{BT01,LLMWZ99,LMW02a,LMW02b,PS00} and the references cited 
therein for more details and further applications).
Two core problems in this context are 
{\sc Closest Substring}~\cite{LMW02b} and 
{\sc Consensus Patterns}~\cite{LMW02a}:
\begin{center}
\begin{minipage}[t]{15cm}
{\bf Input:} $k$ strings $s_1, s_2,\dots , s_k$ over alphabet~$\Sigma$
and non-negative integers~$d$ and~$L$.\\
~\\
{\bf Question in case of {\sc Closest Substring}:} 
Is there a string~$s$ of length~$L$, and for $i=1,\ldots,k$, a substring $s^{\prime}_i$ of length~$L$ such that, for all
$i=1,\dots , k$,  $d_H(s,s^{\prime}_i)\leq d$? 
(Here $d_H(s,s_i')$ denotes the Hamming distance between~$s$ and~$s_i'$.)\\
~\\
{\bf Question in case of {\sc Consensus Patterns}:} 
Is there a string~$s$ of length~$L$, 
and for $i=1,\ldots,k$, a substring $s^{\prime}_i$ of length~$L$ such that,
$\sum_{i=1}^{k} d_H(s,s^{\prime}_i)\leq d$? \\ 
\end{minipage}
\end{center}

What is currently known about these two problems is summarized as follows.

{\bf The {\sc Closest Substring} Problem.}
\begin{enumerate}
\item {\sc Closest Substring\/} is $\NP$-complete, and remains so for the
special case of  the {\sc Closest String\/} problem, where the
string~$s$ that we search for is  of same length as the input
strings. {\sc Closest String\/} is $\NP$-complete even for the further
restriction to a binary alphabet~\cite{FL97,LLMWZ99}.
\item On the positive side, both {\sc Closest Substring} and {\sc Closest String} admit polynomial
time approximation schemes (PTAS's), where the objective function is the maximum length of the
string $s$ ~\cite{LMW99,LMW02a,LMW02b,Ma00}.
\item  In the PTAS's for both {\sc Closest String} and {\sc Closest Substring}, the exponent
of the polynomial bounding the running time depends on the goodness of the
approximation.  These are not efficient PTAS's (EPTAS's) 
in the sense of \cite{CT97} and therefore are probably not useful for
bioinformatics practice.  Whether EPTAS's are possible for these approximation problems, or whether they are
$W[1]$-hard (for the parameter $k =1/\epsilon$, where the approximation is to within a factor of $(1+\epsilon)$
of optimal), currently remains open.
\item
{\sc Closest String} is {\em fixed-parameter tractable\/} with respect
to the parameter~$d$, and
can be solved in time $O(kL+kd\cdot d^d)$~\cite{GNR01}.
\item
{\sc Closest String} is also fixed-parameter tractable
with respect to the parameter~$k$, but here the 
exponential parametric function is much faster growing,
and the algorithm is probably of less practical use
(see, however, \cite{GHN02} for some encouraging experimental results
also in this case). 
\end{enumerate}

{\bf The {\sc Consensus Patterns} Problem.}
\begin{enumerate}
\item
{\sc Consensus Patterns} is $\NP$-complete and remains so for the restriction to a binary alphabet~\cite{LMW99}.
\item 
{\sc Consensus Patterns} admits a PTAS~\cite{LMW99,LMW02a}, where the objective function is the maximum length
of the string $s$.
\item
The known PTAS's for {\sc Consensus Patterns} are not EPTAS's, and whether EPTAS's are possible, or whether
PTAS approximation for this objective function is $W[1]$-hard, is an important issue that also currently remains open.
\end{enumerate}

The key distinguishing point between {\sc Closest Substring}
and {\sc Consensus Patterns} lies in the definition of the distance 
measure $d$ between the ``solution'' string $s$ and the substrings of the $k$ input strings.
Whereas {\sc Closest Substring} uses a maximum distance metric, 
{\sc Consensus Patterns} uses the sum of distances metric.
This is of particular importance
when discussing values of parameter~$d$ occurring in practice.
Whereas it makes good sense for many applications 
to assume that $d$ is a fairly small number in case of {\sc Closest
Substring}, this is much 
less reasonable in the case of {\sc Consensus Patterns}.
This will be of some importance when discussing our result 
for {\sc Consensus Patterns}.

Many algorithms applied in practice try to solve motif search
problems exactly, often using enumerative approaches 
in combination with heuristics~\cite{BST02,BT01,PS00}.
In this paper, we explore the parameterized complexity of the basic motif problems in the framework of
\cite{DF99}.

{\bf Our Main Results.}

Unfortunately, our main results are negative ones: we
show that {\sc Closest Substring}
and {\sc Consensus Patterns} are W[1]-hard with respect to the parameter~$k$
of the number of input strings,
even in case of a binary alphabet.

 For unbounded alphabet size, we show that the problems are
$\W[1]$-hard for the combined parameters $L$, $d$, and~$k$.
In the case
of constant alphabet size, the complexity of the problems remain open
when parameterized by $d$ and~$k$ together, or by~$d$ alone.
Note that in the case
of {\sc Consensus Patterns\/} our result gains particular importance,
because here the distance parameter~$d$ usually is not small,
whereas assuming that $k$ is small is reasonable.
Until now, it was known only that if one additionally considers
the substring length~$L$ as a parameter, then running times
exponential in~$L$ can be achieved~\cite{BST02,EW01,Sag98}. An
overview on known parameterized complexity results for {\sc Closest
Substring\/} and {\sc Consensus Patterns\/} is given in
Table~$\ref{restable}$.
\begin{table}
\begin{center}
\begin{tabular}[t]{|c|c|c|}
\hline
parameter & constant size alphabet & unbounded alphabet\\
\hline
$d$ & ? & $\W[1]$-hard$^{(*)}$\\
$k$ & $\W[1]$-hard$^{(*)}$ & $\W[1]$-hard$^{(*)}$\\
$d,k$ & ? & $\W[1]$-hard$^{(*)}$\\
$L$ & $\FPT$ & $\W[1]$-hard$^{(*)}$\\
$d,k,L$ & $\FPT$ & $\W[1]$-hard$^{(*)}$\\
\hline
\end{tabular}
\end{center}
\caption{Overview on the parameterized complexity of $\CSS$ and {\sc
Consensus Patterns} with respect to different parameterizations, where
$k$ is the number of given strings, $L$ is the
  length of the substrings we search for, and $d$ is the Hamming
  distance allowed. Results from this paper are marked by $(*)$. The $\FPT$
  results for constant size alphabet can be achieved by enumerating all
  length~$L$ strings over $\Sigma$. Open questions are indicated by a question mark.}
\label{restable}
\end{table}

We achieve our results by giving parameterized many-one reductions
from the W[1]-complete {\sc Clique} problem to the respective problems.
It is important here to note that parameterized reductions 
are much more fine-grained than conventional polynomial time 
reductions used in NP-completeness proofs, since
parameterized reductions have to take care of the parameters.
Establishing that {\sc Closest Substring} and 
{\sc Consensus Patterns} are W[1]-hard with respect to the parameter~$k$
requires significantly more technical effort than the
already known demonstrations of NP-completeness.
Finally, our work gives strong theory-based
support for the common intuition that {\sc Closest Substring\/}
($\W[1]$-hard) seems to be a much harder problem than
{\sc Closest String\/} (in $\FPT$~\cite{GNR01}).
Notably, this could {\em not\/} be expressed by ``classical
complexity measures,'' since both problems are
$\NP$-complete as well as both do have a PTAS.

Our work is organized as follows.
In Section~\ref{prelim},
we provide some background on parameterized complexity theory 
and we give a brief overview on related computational biology results.
Afterwards, in Section~\ref{par_d_and_k}, we present a parameterized
reduction of {\sc Clique} to {\sc Closest Substring} 
in case of unbounded input alphabet size.
Then, in Section~\ref{par_k}, this is specialized to the case 
of binary input alphabet.
Finally, Section~\ref{mcpsec} gives similar constructions and results
for {\sc Consensus Patterns} and the paper concludes with a brief 
summary and open questions in Section~\ref{concl}.

\section{Preliminaries and Previous Work}\label{prelim}
In this section, we start with a brief introduction
to parameterized complexity (more details can be found 
in the monograph~\cite{DF99} and the recent survey
articles~\cite{AGN01,Fel01}).  

\subsection{A Crash Course in Parameterized Complexity}
Given an undirected graph $G=(V,E)$ with vertex set~$V$,
edge set~$E$, and a positive integer~$k$,
the $\NP$-complete {\sc Vertex Cover\/}
problem is to determine
whether there is a subset of vertices $C\subseteq V$ with
$k$ or fewer vertices such that
each edge in~$E$ has at least one of its endpoints in~$C$.
{\sc Vertex Cover\/} is {\em fixed-parameter tractable\/}.
There now are algorithms solving it in time less than
$O(kn + 1.3^k)$~\cite{CKJ01,NR99}. The corresponding complexity class
is called~$\FPT$.
By way of contrast, consider the $\NP$-complete {\sc Clique\/}
problem:
Given an undirected graph $G=(V,E)$ and a positive integer~$k$,
{\sc Clique\/} asks whether there is a
subset of vertices $C\subseteq V$
with at least $k$ vertices
such that $C$ forms a clique by having all possible edges
between the vertices in~$C$.
{\sc Clique\/} appears to be {\em fixed-parameter intractable\/}:
It is {\em not\/} known whether it can be solved in time $f(k)\cdot
n^{O(1)}$, where
$f$ might be an arbitrarily fast growing function only depending
on~$k$.

The best known algorithm solving {\sc Clique\/} runs in time
$O(n^{ck/3})$~\cite{NP85}
, where $c$ is the exponent in the time
bound for multiplying two integer $n\times n$ matrices
(currently best known, $c=2.38$, see~\cite{CW90}).
The decisive point is that $k$ appears in the exponent of~$n$,
and there seems to be no way ``to shift the combinatorial
explosion only into~$k$,'' independent
from~$n$.

Downey and Fellows developed a
completeness program for showing parameterized intractability~\cite{DF99}.
However, the completeness theory of parameterized intractability
involves significantly more technical effort
(as will also become clear when following the proofs presented
in this paper). We very briefly sketch some integral
parts of this theory in the following.

Let $L,L'\subseteq \Sigma^*\times \nat$ be two parameterized
languages.\footnote{In general, the second component (representing
the parameter) can also be drawn from $\Sigma^*$; for most cases, and,
in particular, in this paper, assuming the parameter to be a positive
integer is sufficient.}
For example, in the case of {\sc Clique\/}, the first component is
the input graph
coded over some alphabet~$\Sigma$ and the second component is the
positive integer~$k$, that is, the parameter.
We say that {\em $L$ reduces to $L'$ by a standard
parameterized $m$-reduction\/} if there are functions $k\mapsto k'$
and $k\mapsto k''$ from $\nat$ to $\nat$ and a function
$(x,k)\mapsto x'$ from $\Sigma^*\times\nat$ to $\Sigma^*$ such that
\begin{enumerate}
\item $(x,k)\mapsto x'$ is computable in time
$k'' |x|^c$ for some constant~$c$ and
\item $(x,k)\in L$ iff $(x',k')\in L'$.
\end{enumerate}

Notably, most reductions from classical complexity turn out
{\em not\/} to be parameterized ones.
The basic reference degree for parameterized intractability,
$\W[1]$, can be defined as the class of parameterized languages that are equivalent to 
the {\sc Short Turing Machine Acceptance\/}
problem (also known as the {\sc $k$-Step Halting\/} problem).
Here, we want to determine, for an input consisting of a
nondeterministic Turing machine~$M$ (with unbounded nondeterminism and
alphabet size), and a string~$x$, whether~$M$ has a computation path
accepting~$x$ in at most $k$~steps.
This can trivially be solved in time $O(n^{k+1})$ by exploring all
$k$-step computation paths exhaustively, and we would be surprised if
this can be much improved.

Therefore, this is
the parameterized analogue of the {\sc Turing Machine Acceptance\/}
problem that is the basic generic $\NP$-complete problem in classical
complexity theory, and the conjecture that $\FPT\not= \W[1]$ is very much
analogous to the conjecture that $\P\not= \NP$.  Other problems that
are {\em $\W[1]$-complete\/} (there are many) include {\sc Clique\/}
and {\sc Independent Set\/}, where the parameter is the size of the
relevant vertex set \cite{DF95,DF99}.

{F}rom a practical point of view, $\W[1]$-hardness 
gives a concrete indication that a parameterized problem with parameter $k$
problem is unlikely to allow for an algorithm with a running time of
the form $f(k)\cdot n^{O(1)}$.

\subsection{Motivation and Previous Results}
Many biological problems with respect to DNA, RNA, or protein sequences
can be solved based on consensus word 
analysis~\cite[Section~8.6]{Pev00}; {\sc Closest Substring\/} and {\sc
Consensus Patterns\/} are central problems in this context~\cite{GJL99,LLMWZ99,LMW02a,LMW02b}.
Applications include locating binding sites and finding conserved
regions in unaligned sequences for genetic drug target identification,
for designing genetic probes, and for universal PCR primer design.
These problems can be regarded as various generalizations of the
common substring problem, allowing
errors (see~\cite{LLMWZ99,LMW99,LMW02a,LMW02b} and references there).
This leads to {\sc Closest Substring\/} and {\sc Consensus Patterns\/},
where errors are modeled by the (Hamming) distance parameter~$d$.

There is a straightforward factor-2-approximation algorithm
for {\sc Closest Substring\/}.
The first better-than-2 approximation with factor
$2-2/(2|\Sigma |+1)$ was given by Li {\em et al.\/}~\cite{LMW99}.
Finally, there are PTASs for {\sc Consensus Patterns\/}~\cite{LMW99,LMW02a}
as well as for {\sc Closest Substring\/}~\cite{LMW02b,Ma00}, both of which,
however, have impractical running times.

Concerning exact (parameterized) algorithms, we only briefly
mention that, e.g., Sagot~\cite{Sag98} studies motif discovery by
solving {\sc Closest Substring\/}, Evans and Wareham~\cite{EW01} give $\FPT$
algorithms for the same problem, and Blanchette {\em et al.}~\cite{BST02} developed
a so-called phylogenetic footprinting method for a slightly more
general version of {\sc Consensus Patterns\/}. All these results,
however, make essential use of the parameter ``substring length''~$L$
and the running times show exponential behavior with respect
to~$L$. To circumvent the
computational limitations for larger values of~$L$, many heuristics
were proposed, e.g., Pevzner
and Sze~\cite{PS00} present algorithms called WINNOWER (wrt.\ {\sc
Closest Substring\/}) and SP-STAR (wrt.\ {\sc Consensus Patterns\/}),
and Buhler and Tompa~\cite{BT01} use random projections to find
closest substrings. Our analysis makes a first step towards showing 
that, for exact solutions, we have to include~$L$ in the exponential growth;
namely, we show that it is highly unlikely to find algorithms
with a running time exponential {\em only\/} in~$k$.

\section{{\sc Closest Substring\/}: Unbounded Alphabet}
\label{par_d_and_k}

We first describe a reduction from the $\W[1]$-hard $\CLIQUE$ problem to $\CSS$
which is a parameterized $m$-reduction with respect to the aggregate parameter $(L, d, k)$ in case of
unbounded alphabet size.

\subsection{Reduction of $\CLIQUE$ to $\CSS$}
\label{reductiondandk}

A $\CLIQUE$ instance is given by an undirected graph~$G=(V,E)$, with a
set~$V=\{v_1, v_2,\dots , v_n\}$ of $n$~vertices, a set~$E$ of $m$~edges, and
a positive
integer~$k$ denoting the desired clique size. We describe how to
generate a set~$S$ of $k\choose 2$ strings such that~$G$ has a clique
of size~$k$ iff there is a string~$s$ of length~$L:=k+1$ such that every
$s_i\in S$ has a substring~$s_i^{\prime}$ of length~$L$ with $d_H(s,
s_i^{\prime})\leq d:=k-2$. If a string~$s_i\in S$ has a
substring~$s_i^{\prime}$ of length~$L$ with $d_H(s, s_i^{\prime})\leq
d$, we call~$s_i^{\prime}$ a {\em match}. We assume $k>2$, because
$k=1,2$ are trivial cases.

\noindent{\bf Alphabet.} The alphabet of the produced instance is
given by the disjoint union of the following sets:
\begin{itemize}
\item $\{\,\sigma_i \mid v_i\in V\,\}$, i.e., an alphabet
symbol for every vertex of the input graph; we call them {\em
encoding symbols};
\item $\{\, \varphi_j\mid j=1,\dots ,\mbox{${k\choose 2}$}\,\}$, i.e.,
a unique symbol for every of the ${k\choose 2}$~produced strings; we
call them {\it string identification symbols};
\item $\{\#\}$ which we call the {\em synchronizing symbol}.
\end{itemize}
This makes a total of $n+{k \choose 2}+1$ alphabet symbols.

\noindent{\bf Choice strings.}
We generate a set of $k \choose 2$ {\em choice strings}
$S_c=\{c_{1,2}, \dots , c_{1,k}$, $c_{2,3}$, $c_{2,4}, \dots , c_{k-1,k}\}$
and we assume that the strings in~$S_c$ are ordered as shown. {\em Every}
choice string will
encode the whole graph; it consists of $m$ concatenated strings, each of
length~$k+1$, called {\em blocks}; by this, we have one block for every
edge of the graph. The blocks will be separated by {\em barriers},
which are length~$k$ strings consisting of~$k$ identification
symbols corresponding to the respective string. A choice
string~$c_{i,j}$, which, according to the given order, is the $i^{\prime}$th
choice string in $S_c$,
is given by
\[
\begin{array}{l}
c_{i,j}:=
\langle\mbox{block}(i,j,e_1)\rangle\,(\varphi_{i^{\prime}})^k\,
\langle\mbox{block}(i,j,e_2)\rangle\,(\varphi_{i^{\prime}})^k
\dots
(\varphi_{i^{\prime}})^k\,\langle\mbox{block}(i,j,e_m)\rangle,
\end{array}
\]
where $e_1, e_2,\dots , e_m$ are the edges of~$G$ and
$\langle\mbox{block}()\rangle$ will be defined below.
The solution string~$s$ will have length~$k+1$, which is exactly the
length of one block.

\noindent{\bf Block in a choice string.} Every block is a string of
length~$k+1$ and it encodes an edge of the input graph. Every choice string
contains a block for every edge of the input graph; different choice strings,
however, encode the edges in different positions of their blocks: For a
block in choice string~$c_{i,j}$, positions~$i$ and~$j$ are called
{\em active} and these positions encode the edge.
Let $e$ be the edge to be encoded
and let~$e$ connect vertices~$v_r$ and~$v_s$, $1\leq r<s\leq n$.
Then, the $i$th position of the block is~$\sigma_r$ in order to
encode~$v_r$ and the $j$th position is~$\sigma_s$ in order to
encode~$v_s$. The
last position of a block is set to the synchronizing symbol~$\#$. Let
$c_{i,j}$ be the $i^{\prime}$th choice string in $S_c$; then,
all remaining positions in the block are set to
$c_{i,j}$'s identification symbol~$\varphi_{i^{\prime}}$.
Thus, the block is given by
\[
\begin{array}{ll}
\langle\mbox{block}(i,j,(v_r,v_s))\rangle\ :=\ 
(\varphi_{i^{\prime}})^{i-1}\,\sigma_r\,(\varphi_{i^{\prime}})^{j-i-1}\,\sigma_s\,(\varphi_{i^{\prime}})^{k-j}\,\#.
\end{array}
\]

\noindent{\bf Values for~$L$ and~$d$.}
We set~$L:=k+1$ and~$d:=k-2$.

\begin{figure}[t]
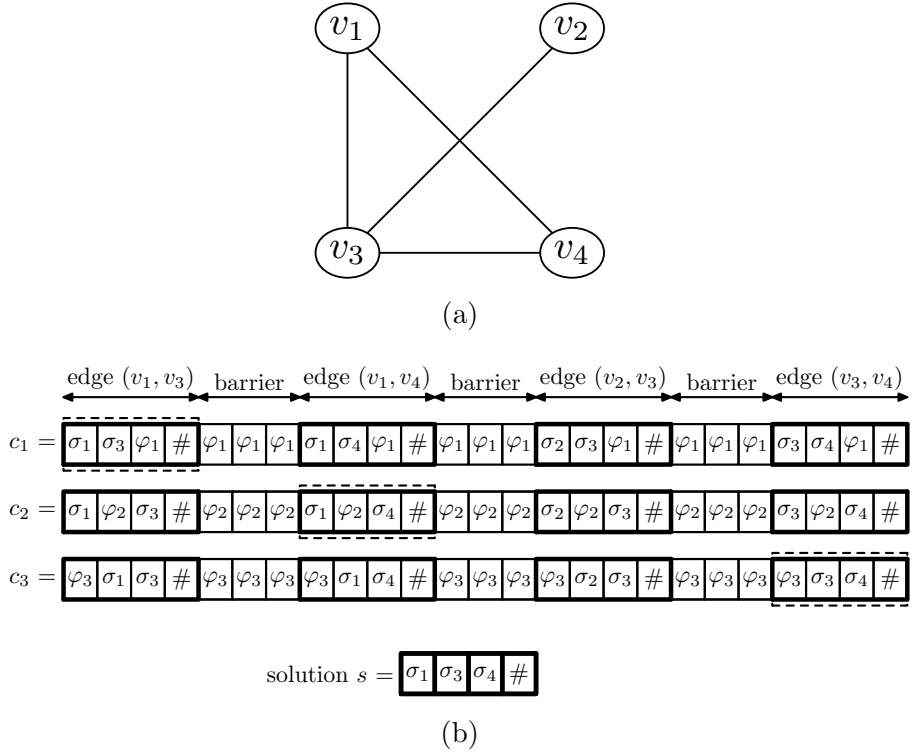

\begin{center}
\epsfig{file=example.1,width=3.9cm}

\smallskip

(a)

\bigskip

\epsfig{file=example.2, width=12cm}

\smallskip

(b)
\end{center}
\caption{Example for the reduction from a $\CLIQUE$ instance~$G$ with
$k=3$ (shown in (a)) to a $\CSS$ instance with bounded alphabet (shown
in (b)) as explained in Example~\ref{redex1}. In~(b), we display
the constructed strings~$c_1$, $c_2$, and~$c_3$
(the contained blocks are highlighted by bold boxes) and the
solution string~$s$ that is found, since $G$ has a clique of size~$k=3$;
$s$~is a string of length $k+1=4$ such that $c_1$,
$c_2$, and~$c_3$ have length~$4$ substrings (indicated by dashed
boxes) that have Hamming distance at most~$k-2=1$ to~$s$.}
\label{exfig1}
\end{figure}

\begin{exmp}
\label{redex1}\rm
Let $G=(V,E)$ be an undirected graph with $V=\{v_1,v_2,v_3,v_4\}$ and
$E=\{(v_1,v_3)$, $(v_1,v_4)$, $(v_2,v_3)$, $(v_3,v_4)\}$ (as
shown in Fig.~$\ref{exfig1}$(a)) and let $k=3$. Using~$G$,
we exhibit the above construction of ${k\choose 2}=3$ choice strings~$c_1$,
$c_2$, and $c_3$ (as shown in Fig.~$\ref{exfig1}$(b)). Note that, in
the described construction, the strings were called $c_{1,2}$, $c_{1,3}$, and
$c_{2,3}$, but, here, for the ease of presentation, we call them~$c_1$,
$c_2$, and~$c_3$. We claim that (which will be proven in the following
subsection) there exists a clique of size~$k$ in~$G$ iff
there is a string~$s$ of length $L:={k\choose 2}+1=4$ such that, for
$i=1, 2, 3$, each~$c_i$ contains a length~$4$ substring~$s_i$ with $d_H(c_i,
s_i)\leq d:=k-2=1$. 

The choice strings are over an alphabet consisting of $\{\sigma_1, \sigma_2,
\sigma_3, \sigma_4\}$ (the encoding symbols, i.e., one symbol for
every node of~$G$), $\{\varphi_1, \varphi_2, \varphi_3\}$ (the string
identification symbols),
and $\{\#\}$ (the synchronizing symbol). Every string~$c_i$,
$i=1, 2, 3$ consists of four blocks, each of which encodes an edge
of the graph. Every block is of length~${k\choose 2}+1=4$ and has $\#$
at its last position. The blocks are separated by barriers consisting
of $(\varphi_i)^k=(\varphi_i)^3$.

In string~$c_1$, positions~$1$ and~$2$ within a block are active and
encode the corresponding edge (in~$c_2$ positions~$1$ and~$3$, and,
in~$c_3$ positions~$2$ and~$3$ within a block are active). All
of the first $k$ positions of a block in
string~$c_i$, $i=1, 2, 3$ which are not active, contain the
$\varphi_i$ symbol. Thus, e.g., the block in~$c_1$ encoding the
edge~$(v_1,v_3)$ is given by $\sigma_1\sigma_3\varphi_1\#$. Further
details can be found in Fig.~$\ref{exfig1}$.

The closest substring that corresponds to the $k$-clique in~$G$ consisting of
vertices~$v_1$, $v_3$, and~$v_4$ is~$\sigma_1\sigma_3\sigma_4\#$. The
corresponding matches are
$\sigma_1\sigma_3\varphi_1\#$ in~$c_1$ (encoding the edge~$(v_1,
v_3)$), $\sigma_1\varphi_2\sigma_4\#$ in~$c_2$ (encoding the edge~$(v_1,
v_4)$), and $\varphi_3\sigma_3\sigma_4\#$ in~$c_3$ (encoding the edge~$(v_3,
v_4)$).
\end{exmp}

\subsection{Correctness of the Reduction}
\label{correctnessdandk}

To prove the correctness of the proposed reduction, we have to show an
equivalence, consisting of two directions. The easier one is to see that a
$k$-clique implies a closest substring fulfilling the given requirements.


\begin{prop}
\label{onedirdandk}
For a graph with a $k$-clique, the construction in
Subsection~$\ref{reductiondandk}$ produces an instance of
$\CSS$ which has a solution, i.e., there is a string~$s$ of length~$L$ such
that every $c_{i,j}\in S_c$ has a substring~$s_{i,j}$ with
$d_H(s,s_{i,j})\leq d$.
\end{prop}
\begin{proof}
Let the input graph have a clique of size~$k$. Let $h_1, h_2,\dots ,
h_k$ denote the indices of the clique's vertices, $1\leq h_1<h_2<\dots
<h_k\leq n$. Then, we claim that a solution for the produced $\CSS$ instance is
$$s:=\sigma_{h_1}\sigma_{h_2}\dots\sigma_{h_k}\#.$$ Consider
choice string~$c_{i,j}$, $1\leq i<j\leq k$. As the vertices $v_{h_1},
v_{h_2},\dots , v_{h_k}$ form a clique, we have an edge connecting~$v_{h_i}$
and~$v_{h_j}$. 
Choice string $c_{i,j}$ contains a
block~$s_{i,j}:=\langle\mbox{block}(i,j,(v_{h_i},v_{h_j}))\rangle$
encoding this edge:
\[
\begin{array}{l}
s_{i,j}\ :=
(\varphi_{i^{\prime}})^{i-1}\,\sigma_{h_i}\,(\varphi_{i^{\prime}})^{j-i-1}\,\sigma_{h_j}\,(\varphi_{i^{\prime}})^{k-j}\#,
\end{array}
\] 
where $i^{\prime}$ is the number (according to the given order) of the choice
string in~$S_c$. We have $d_H(s,s_{i,j})=k-2$, and we can
find such a block for every~$c_{i,j}$, $1\leq i<j\leq k$.
\end{proof}

For the reverse direction, we show in Proposition~$\ref{reversedandk}$
that a solution in the produced {\sc Closest Substring\/} instance
implies a $k$-clique in the input graph. For this, we need the
following two lemmas, which show that a solution to the
instance constructed in Subsection~$\ref{reductiondandk}$ has encoding
symbols at its first~$k$ positions and the synchronizing symbol~$\#$ at its
last position.

\begin{lem} \label{dandklem1} A closest substring~$s$ contains at least two
encoding symbols and at least one synchronization symbol.
\end{lem}

\begin{proof}
Let $s$ be a solution of the {\sc Closest Substring\/} instance
produced by the construction in Subsection~$\ref{reductiondandk}$. Let
$A_\varphi(s)$ be the set of string identification symbols from
$\{\,\varphi_i\mid 1\leq i\leq \mbox{$k\choose 2$}\,\}$ that occur in~$s$. Let
$S_\varphi(s)\subseteq S_c$ be the subset of choice strings that do
{\em not} contain a symbol from~$A_\varphi(s)$. 

Since~$s$ is of length~$k+1$, we have $|A_\varphi(s)|\leq
k+1$. Therefore, for $k\geq 4$, there are at least ${k\choose 2}-(k+1)$ choice
strings in~$S_\varphi(s)$. We show that with less than two
encoding symbols and no synchronizing symbol, we cannot find
matches for~$s$ (with maximally allowed Hamming distance~$d=k-2$) in the
choice strings of~$S_\varphi(s)$. Observe that, in every choice
string, because of the barriers, every length~$k+1$ substring contains at most
two encoding
symbols and at most one symbol~$\#$. Observe further that, taken a
choice string from~$S_\varphi(s)$, positions with symbols from
$\{\,\varphi_i\mid 1\leq i\leq \mbox{$k\choose 2$}\,\}$ cannot
coincide with the corresponding positions in~$s$. Therefore, $s$ has a match
in such a string only if~$s$ has two encoding symbols and one symbol~$\#$
that all coincide with the corresponding positions in the selected
substring. This proves the claim for $k\geq 4$. Regarding $k=3$, if
$|A_\varphi(s)|<3$, then the above argument applies here, too. If, however,
$|A_\varphi(s)|=3$, a length $4$ substring in every choice string has at least
two positions that do not coincide with the corresponding positions in~$s$.
\end{proof}

Based on Lemma~\ref{dandklem1}, we can now exactly
specify the numbers and positions of the encoding and synchronizing symbols in
the closest substring.

\begin{lem} \label{dandklem2} A closest substring~$s$ contains encoding symbols
at its first $k$~positions and a symbol~$\#$ at its last position.
\end{lem}

\begin{proof}
Let~$n_{\#}(s)$ denote the number of symbols~$\#$ in~$s$,
let $n_{\varphi}(s)$ denote the number of string identification
symbols in~$s$, and let $n_{\sigma}(s)$ denote the number of encoding
symbols in~$s$. Let $S_\varphi(s)\subseteq S_c$ be the subset of choice
strings whose string identification symbol does not occur in~$s$.
In the following, we establish a lower bound on the number of strings
in~$S_\varphi(s)$ and an upper bound on the number of strings
from~$S_\varphi(s)$ in which we can find a match for~$s$. Comparing
these bounds, we will show that, if $n_{\#}(s)>1$, then there are
choice strings in~$S_\varphi(s)$ in which we cannot find a match;
we will conclude that $n_{\#}(s)=1$. Then, we will show that, if
$n_{\sigma}(s)<k$, then again there are strings
in~$S_\varphi(s)$ without a match; we will conclude that $n_{\sigma}(s)=k$.

Regarding the size of~$S_\varphi(s)$, a lower bound on its size
is~$|S_\varphi(s)|\geq {k\choose 2}-n_{\varphi}(s)$. To explain the upper
bound on the number of strings from~$S_\varphi(s)$ in which we can find a
match for~$s$, we recall that such matches must contain two encoding
symbols and one symbol~$\#$ that all coincide with the corresponding
positions in~$s$. On the one hand, the synchronizing symbol of a block must
coincide with a symbol~$\#$ in~$s$. On the other hand, in all blocks of a
choice string, its encoding symbols are in fixed positions relative to the
block's synchronizing symbol, e.g., in choice string~$c_{1,2}$, the encoding
symbols are located only at the first and second position and $\#$ at the last
position of a block in~$c_{1,2}$. For these two reasons, one symbol~$\#$
in~$s$ can provide matches in at most $n_{\sigma}(s)\choose 2$ choice strings
from~$S_\varphi(s)$. Consequently, $n_{\#}(s)$ many symbols~$\#$ in~$s$ can
provide matches in at most $n_{\#}(s)\cdot{n_{\sigma}(s)\choose 2}$ choice
strings from~$S_\varphi(s)$.

Summarizing, we have at least ${k\choose 2}-n_{\varphi}(s)$ choice strings
in~$S_\varphi(s)$ and we can find matches in at
most~$n_{\#}(s)\cdot{n_{\sigma}(s)\choose 2}$ many of them. Thus, we find
matches for~$s$ in all choice strings only if 
\begin{equation}\label{dandklem2inq}
n_{\#}(s)\cdot{n_{\sigma}(s)\choose 2}\geq {k\choose 2}-n_{\varphi}(s).
\end{equation}
In order to show that~$s$ contains exactly one
synchronizing symbol, we assume that $n_{\#}(s)>1$ (we
know that $n_{\varphi}(s)\geq 1$ by Lemma~$\ref{dandklem1}$) while~$k>2$, and
show that inequality~$\ref{dandklem2inq}$ is violated.

We know that $k+1=n_{\sigma}(s)+n_{\varphi}(s)+n_{\#}(s)$ and,
by Lemma~$\ref{dandklem1}$, that $n_{\sigma}(s)\geq 2$. Using these, we
conclude, on the one hand, that $n_{\#}(s)\cdot{n_{\sigma}(s)\choose 2}\leq
n_{\#}(s)\cdot{{k+1-n_{\#}(s)}\choose 2}$ and, since $n_{\#}(s)\geq
2$, that
$n_{\#}(s)\cdot{{k+1-n_{\#}(s)}\choose 2}\leq 2\cdot{{k-1}\choose 2}$.
On the other hand, we have that
${k\choose 2}-n_{\varphi}(s)\geq {k\choose 2}-(k-1-n_{\#}(s))$ and, since
$n_{\#}(s)\geq 2$, ${k\choose 2}-(k-1-n_{\#}(s))\geq {k\choose 2}-(k-3)$. For
$k\geq 3$, however we have ${k\choose 2}-(k-3)>2\cdot{{k-1}\choose 2}$.
Thus,  
\[
n_{\#}(s)\cdot{n_{\sigma}(s)\choose 2}\leq
n_{\#}(s)\cdot{{k+1-n_{\#}(s)}\choose 2}<{k\choose
2}-(k-1-n_{\#}(s))\leq {k\choose 2}-n_{\varphi}(s),
\]
i.e.,
there are choice strings in~$S_\varphi(s)$ which contain no match
for~$s$, a contradiction. Since (Lemma~$\ref{dandklem1}$) $n_{\#}(s)\geq 1$,
we conclude that~$n_{\#}(s)=1$.

In order to show that~$s$ contains exactly~$k$ encoding symbols, 
we assume that $n_{\sigma}(s)<k$ while~$k>2$ and~$n_{\#}(s)=1$, and show that
inequality~$\ref{dandklem2inq}$ is violated. Since
$k+1=n_{\sigma}(s)+n_{\varphi}(s)+n_{\#}(s)=n_{\sigma}(s)+n_{\varphi}(s)+1$,
we have ${k\choose 2}-n_{\varphi}(s)={k\choose 2}-(k-n_{\sigma}(s))$ and, thus,
\[
{n_{\sigma}(s)\choose 2}<{k\choose
2}-(k-n_{\sigma}(s))\leq {k\choose 2}-n_{\varphi}(s),
\]
i.e., again, some strings in~$S_\varphi(s)$ have no match for~$s$, a
contradiction. Thus, on the one hand, we have $n_{\sigma}(s)\geq
k$, and, on the other hand, we have $n_{\#}(s)=1$ and, therefore,
$n_{\sigma}(s)\leq k$.

Note that, if an encoding symbol is located {\em after} the
synchronizing symbol in~$s$, then, due to the barriers, it is not
possible that both~$\#$ and this encoding symbol coincide
with the respective positions in a choice string
from~$S_\varphi(s)$. Therefore, symbol~$\#$ is located at the last position
of~$s$.
\end{proof}



\begin{prop}
\label{reversedandk}
The first~$k$ characters of a closest substring correspond to $k$~vertices of
a clique in the input graph.
\end{prop}

\begin{proof}
By Lemma~$\ref{dandklem2}$, a
closest substring~$s$ has encoding symbols at its first $k$ positions and a
synchronizing symbol at its last position. Consequently, the blocks are
the only possible matches of~$s$ in the choice string. Now,
assume that $s=\sigma_{h_1}\sigma_{h_2}\dots\sigma_{h_k}\#$ for $h_1,h_2,\dots
,h_k\in\{1,\dots ,n\}$. Consider any two $h_i, h_j$, $1\leq i<j\leq
k$, and choice string $c_{i,j}$. Recall that in this choice string,
the blocks encode edges at their $i$th and $j$th position, they have~$\#$
at their last position, and all their other positions are set to a string
identification symbol unique for this choice string. Thus, we can only find a
block that is a match if there is a block with $\sigma_{h_i}$ at its $i$th
position and $\sigma_{h_j}$ at its $j$th position. We have such a block only
if there is an edge connecting~$v_{h_i}$ and~$v_{h_j}$. Summarizing, the
closest substring~$s$ implies that there is an edge between every pair of
$\{v_{h_1},v_{h_2},\dots ,v_{h_k}\}$; these vertices form a $k$-clique
in the input graph.   
\end{proof}

Propositions~$\ref{onedirdandk}$ and~$\ref{reversedandk}$ establish
the following hardness result.
Note that hardness for the combination of all three parameters also implies
hardness for each subset of the three.

\begin{thm}
$\CSS$ with unbounded alphabet is $\W[1]$-hard for every combination of the
parameters~$L$, $d$, and~$k$.
\end{thm}

\section{{\sc Closest Substring\/}: Binary Alphabet}
\label{par_k}

We modify the reduction from Section~\ref{par_d_and_k}
to achieve a $\CSS$ instance with binary alphabet proving a $\W[1]$-hardness
result also in this case. In contrast to
the previous construction, we cannot encode every vertex
with its own symbol and we cannot use a unique symbol for every produced
string. Also, we have to find new ways to ``synchronize'' the matches of
our solution, a task previously done by the synchronizing
symbol~\#. To overcome these problems, we construct an additional ``complement
string'' for the input instance and we lengthen the blocks in the produced
choice strings considerably.

\subsection{Reduction of $\CLIQUE$ to $\CSS$}
\label{reduction_k}

\noindent{\bf Number strings.} To encode integers between~$1$ and~$n$,
we introduce {\em number strings}
$\langle\mbox{number}(\pos)\rangle$, which have length $n$ and which
have symbol~``1'' at position $\pos$ and symbol~``0''
elsewhere: $0^{pos-1}\,1\,0^{n-pos}$. In
contrast to the reduction from Section~$\ref{par_d_and_k}$, now we
use these number strings to encode the vertices of a graph.

\noindent{\bf Choice strings.} As in Section~$\ref{par_d_and_k}$, we
generate a set of $k \choose 2$ {\em choice strings} 
$S_c=\{c_{1,2}$,$c_{1,3} \dots, c_{k-1,k}\}$. 
Again, every choice string will consist of $m$ {\em blocks}, one block
for every edge of the graph. The choice string~$c_{i,j}$ is given by
\[
\begin{array}{l}
c_{i,j}\ :=\
\langle\mbox{block}(i,j,e_1)\rangle
\langle\mbox{block}(i,j,e_2)\rangle
\dots
\langle\mbox{block}(i,j,e_m)\rangle,
\end{array}
\]
where $e_1, e_2,\dots , e_m$ are the edges of the input graph and
$\langle\mbox{block}()\rangle$ is defined below. The length of
a closest substring will be exactly the length of one block.

\noindent{\bf Block in a choice string.} Every block consists of
a front tag, an encoding part, and a back tag. A block in choice
string~$c_{i,j}$ encodes an edge~$e$; let $e$ be
an edge connecting vertices~$v_r$ and~$v_s$, $1\leq
r<s\leq n$, and let~$c_{i,j}$ be the (according to the given order)
$i^{\prime}$th string in $S_c$. Then, the corresponding block is given by
\[
\langle\mbox{block}(i,j,(v_r,v_s))\rangle\ :=
\langle\mbox{front\_tag}\rangle\langle\mbox{encode}(i,j,(v_r,v_s))\rangle\langle\mbox{back\_tag}(i^{\prime})\rangle.
\]

\noindent{\bf Front tags.} 
We want to enforce that a closest substring
can only match substrings at certain positions in
the produced choice strings, using front tags:
\[
\langle\mbox{front\_tag}\rangle\ :=\ (1^{3nk}0)^{nk},
\]
i.e., a front tag has length~$(3nk+1)\cdot nk$.
By this arrangement, the closest substring~$s$ and every match of~$s$ start
(as will be shown in Subsection~$\ref{correctnessk}$) with the front tag.

\noindent{\bf Encoding part.}
The encoding part consists of $k$~sections, each
of length~$n$. The encoding part corresponds to the blocks
used in Section~$\ref{par_d_and_k}$. As a consequence, in
$\langle\mbox{block}(i,j,e)\rangle$ the $i$th and $j$th section are
called {\em active} and encode edge~$e=(v_r,v_s)$, $1\leq r<s\leq n$;
section~$i$
encodes~$v_r$ by~$\langle\mbox{number}(r)\rangle$ and section~$j$
encodes~$v_s$ by~$\langle\mbox{number}(s)\rangle$. The other sections
except for~$i$ and~$j$ are called {\em inactive} and are given by
$\langle\mbox{inactive}\rangle:=0^n$. Thus,
\[
\langle\mbox{encode}(i,j,(v_r,v_s))\rangle := 
(\langle\mbox{inactive}\rangle)^{i-1}\,\langle\mbox{number}(r)\rangle\,(\langle\mbox{inactive}\rangle)^{j-i-1}\,\langle\mbox{number}(s)\rangle\,(\langle\mbox{inactive}\rangle)^{k-j}.
\]


\noindent{\bf Back tag.}
The back tag of a block is intended to balance the Hamming distance of the
closest substring to a block, as will be explained later. The
back tag consists of $k\choose 2$ sections, each section has
length~$nk-2k+2$. The $i^{\prime}$th section consists of
symbols~``1,'' all other sections consist of symbols~``0'':
\[
\langle\mbox{back\_tag}(i^{\prime})\rangle\ :=\
0^{(i^{\prime}-1)(nk-2k+2)}1^{nk-2k+2}0^{({k\choose
2}-i^{\prime})(nk-2k+2)}
\]

\noindent{\bf Template string.}
The set of choice strings is complemented by one {\em template string}. It
consists,
in analogy to the blocks in the choice strings, of three parts:
A front tag of length~$(3nk+1)\cdot nk$, followed by a length~$nk$
string of symbols~``1,'' followed by a length~${k\choose 2}(nk-2k+2)$ string of
symbols~``0.''
Thus, the template string has the same length as a block in a
choice string, i.e.,~$(3nk+1)\cdot nk+nk+{k\choose 2}(nk-2k+2)$.

\noindent{\bf Values for~$d$ and~$L$.} We set $L:=(3nk+1)\cdot
nk+nk+{k\choose 2}(nk-2k+2)$ and~$d:=nk-k$. As we will show in
Subsection~$\ref{correctnessk}$, the possible matches for a
string of this length are the blocks in the choice strings, and, concerning
the template string, the template string itself.

\noindent{\bf Notation.} For a closest substring~$s$, we denote its
first $(3nk+1)\cdot nk$ symbols (the front tag) by~$s^{\prime}$, the
following $nk$ symbols (its encoding part) by~$s^{\prime\prime}$, and
the last ${k\choose 2}(nk-2k+2)$ symbols (its back tag),
by~$s^{\prime\prime\prime}$. Analogously, the three parts of
the template string~$t$ are denoted~$t^{\prime}, t^{\prime\prime}$,
and $t^{\prime\prime\prime}$. A particular block of a choice
string~$c_{i,j}$, is referred to by~$s_{i,j}$; its three parts are
called $s_{i,j}^{\prime}, s_{i,j}^{\prime\prime}$,
and~$s_{i,j}^{\prime\prime\prime}$.

\begin{figure}
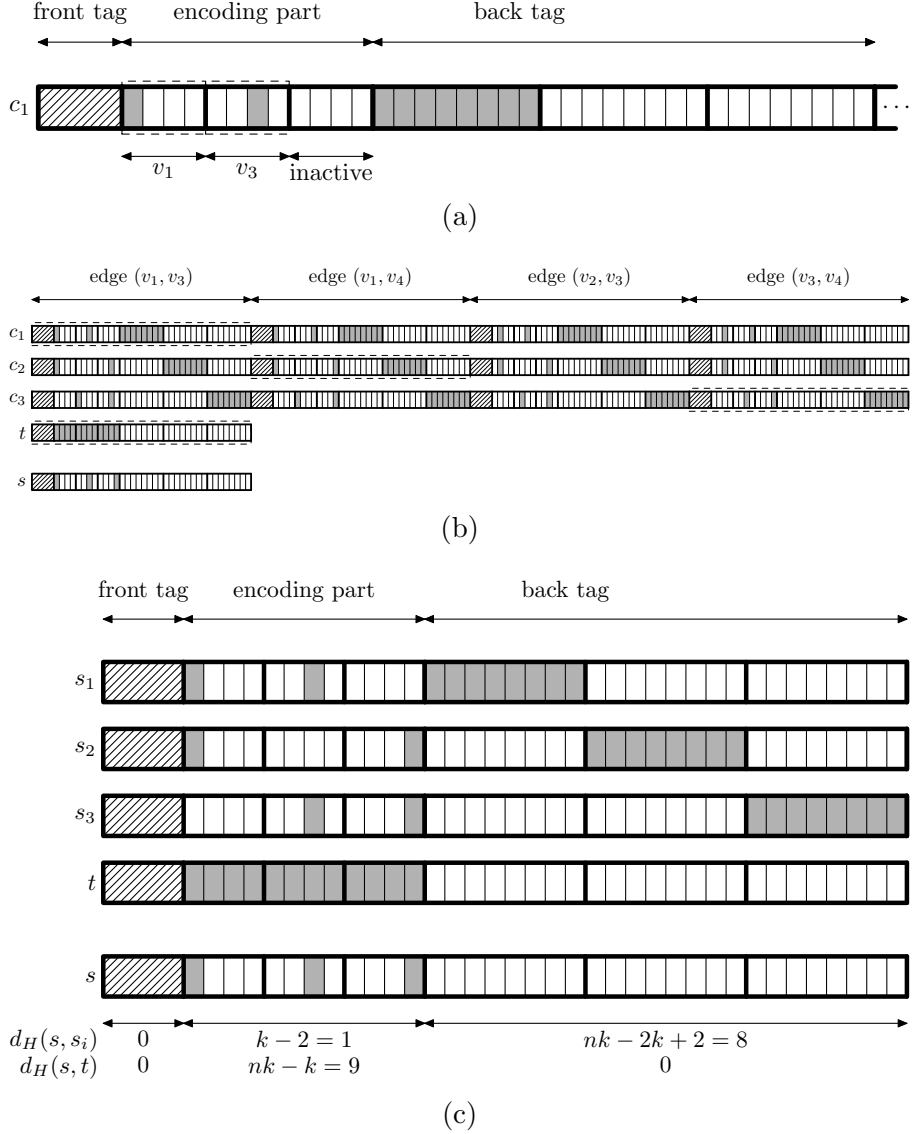

\begin{center}
\epsfig{file=example.5, width=12cm}

\smallskip

(a)

\bigskip

\epsfig{file=example.3, width=12cm}

\smallskip

(b)

\bigskip

\epsfig{file=example.4, width=12cm}

\smallskip

(c)
\end{center}
\caption{Example for the reduction from the $\CLIQUE$ instance~$G$
(shown in Fig.~\ref{redex1}(a)) to a $\CSS$ instance with binary
alphabet as explained in Example~\ref{redex2}. When displaying the
strings, we omit the details of the front tag parts and only indicate
them shortened in their proportion to the other parts of the strings;
all front tag parts in all strings are equal. In the encoding parts and
the back tag parts, we indicate the symbols~``1'' of the construction
by dark boxes, the symbols~``0'' by white boxes. In~(a), we outline the
first block of~$c_1$. In its encoding
part, sections~$1$ and~$2$ (sections are indicated by bold separating
lines) are active (indicated by dashed boxes) and encode the first
edge~$(v_1,v_3)$ of graph~$G$; the remaining third section is
inactive. In its back tag part, the first section is filled with
symbols~``1.'' In~(b), we give an overview on all constructed strings, the
choice strings~$c_1$, $c_2$, and $c_3$, and the template
string~$t$. We also display the closest substring~$s$ that is found,
since $G$ has a clique of size~$k=3$; its matches in $c_1$, $c_2$,
$c_3$, and~$t$ are indicated by dashed boxes. In (c), we focus on
these matches and the solution string~$s$, and state, separately for
the front tag, the encoding, and the back tag part, the Hamming
distances of $s$ to a match~$s_i$, $i=1,2,3$ (the
distances are equal for~$s_1$, $s_2$, and~$s_3$) and to the template
string~$t$.}
\label{exfig2}
\end{figure}

\begin{exmp}\rm
\label{redex2}
Let $G=(V,E)$ be the graph from Example~$\ref{redex1}$, with
$V=\{v_1,v_2,v_3,v_4\}$ and $E=\{(v_1,v_3)$, $(v_1,v_4)$, $(v_2,v_3)$,
$(v_3,v_4)\}$ (as shown in Fig.~$\ref{exfig1}$(a)) and let $k=3$. In
the following, we outline the above construction of ${k\choose
2}=3$ choice strings~$c_1$, $c_2$, and $c_3$ and one template
string~$t$ over alphabet $\Sigma=\{0, 1\}$ as displayed in
Fig.~$\ref{exfig2}$.

Every string $c_1$, $c_2$, and $c_3$ consists of four blocks
corresponding to the four edges of~$G$. Fig.~$\ref{exfig2}$(a)
displays the first block of~$c_1$ corresponding to
edge~$(v_1,v_3)$. It consists of a front tag, an encoding part,
and a back tag. The front tag (not displayed in detail in the figure)
is given by
$\langle\mbox{front\_tag}\rangle:=(1^{3nk}0)^{nk}=(1^{36}0)^{12}$;
all front tags for all blocks in all constructed strings are the same. The
back tag of the first block consists of $k\choose 2$
sections; since the back tag is in the {\em first} string, the {\em
first} section is filled with ``1''s and the remaining sections are
filled with ``0''s. Thus, the back tag is given by
$1^{nk-2k+2}0^{({k\choose 2}-1)(nk-2k+2)}=1^{8}0^{16}$, and all back
tags for blocks in the first string are given like this.
The encoding part consists of $k=3$ sections, each section of
length~$n=4$. In the blocks of string~$c_1$, the first and the second
section are active; in the first block they encode
edge~$(v_1,v_3)$. Therefore, the first section is given by
$\langle\mbox{number}(1)\rangle$ and the second one by
$\langle\mbox{number}(3)\rangle$, the remaining inactive section is
filled with ``0''s. 

Fig.~$\ref{exfig2}$(b) displays an overview on all constructed
strings~$c_1$, $c_2$, $c_3$, and~$t$. In all strings, block~$i$
encodes the $i$th edge, $1\leq i\leq 4$. However, the active sections
of the encoding part and the back tags differ for different
strings. The template string~$t$ consists only of one block, which has
a front tag, a part corresponding to the encoding part, filled with
``1''s, and a part corresponding to the back tag, filled with
``0''s. 

Since~$G$ has a $k$-clique for $k=3$, consisting of vertices~$v_1$, $v_3$,
and~$v_4$, we find a solution~$s$ for the constructed $\CSS$
instance. This~$s$ has a front tag, and its back tag part is
filled with ``0'' symbols. The encoding part encodes the vertices of
the clique, it is given by
$\langle\mbox{number}(1)\rangle\langle\mbox{number}(3)\rangle\langle\mbox{number}(4)\rangle$.

Fig.~$\ref{exfig2}$(c) gives a focus on the matches that are found in
$c_1$, $c_2$, $c_3$, and~$t$, which are, for the choice strings,
referred to by $s_1$, $s_2$, and $s_3$, respectively. The front tag
part~$s'$ has distance $0$ to the front tags $s_1'$, $s_2'$, $s_3'$,
and~$t'$. The encoding part $s''$ contains $k=3$ many~``1''s; $s_1''$,
$s_2''$, $s_3''$ have two ``1''s each and, in each case, these ``1''s
coincide with ``1''s in~$s''$. Therefore, $d_H(s'', s_i)=k-2=1$, $1\leq i\leq
3$. The encoding part of the template string, $t''$, only consists of ``1''s
and, therefore, $d_H(s'',t'')=nk-k$. The back tag $s'''$
only consists of ``0''s; each back tag $s_1'''$, $s_2'''$,
and~$s_3'''$ contains~$nk-2k+2=8$ many
``1''s; therefore~$d_H(s''',s_i''')=8$, $1\leq i\leq
3$. The back tag of the template string, $t'''$, contains only ``0''s and,
hence, $d_H(s''',t''')=0$. Altogether, this shows
that, for $1\leq i\leq 3$, $d_H(s,s_i)=d_H(s,t)=nk-k=9$ as
required.
\end{exmp}

\subsection{Correctness of the Reduction}
\label{correctnessk}


To prove the correctness of the reduction, again the easier direction is to
show that a $k$-clique implies a closest substring fulfilling the given
requirements.

\begin{prop}
\label{onedirk}
For a graph with a $k$-clique, the construction in
Subsection~$\ref{reduction_k}$ produces an instance of
$\CSS$ that has a solution, i.e., there is a string~$s$ of length~$L$ such
that every $c_{i,j}\in S_c$ has a length~$L$ substring~$s_{i,j}$ with
$d_H(s,s_{i,j})\leq d$ and $d_H(s,t)\leq d$.
\end{prop}
\begin{proof}
Let the graph have a clique of size~$k$. Let $h_1, h_2,\dots , h_k$
denote the indices of the clique's vertices, $1\leq h_1<h_2<\dots
<h_k\leq n$. Then, we can find a closest substring~$s$, consisting of three
parts $s'$, $s''$, and $s'''$, as follows: its front tag~$s'$ is given
by $\langle\mbox{front\_tag}\rangle$; its encoding part~$s''$ is given by
$\langle\mbox{number}(h_1)\rangle\langle\mbox{number}(h_2)\rangle\dots
\langle\mbox{number}(h_k)\rangle$; its back tag~$s'''$ is $0^{{k\choose
2}(nk-2k+2)}$. 
It follows from the construction that the choice strings have
substrings that are matches for this~$s$: For every $1\leq i<j\leq k$,
we produced choice string~$c_{i,j}$ with a block~$s_{i,j}$ encoding the edge
between vertices~$v_{h_i}$ and~$v_{h_j}$. For these blocks as well as for
the template string, the following table reports the distance they
have to the solution string, separately for each of their three parts
and in total:
\begin{center}
\begin{tabular}{|c||c|c|c|c|c|}
\hline
 $d_H(\cdot,\cdot)$ & $s^{\prime}$ & $s^{\prime\prime}$ & $s^{\prime\prime\prime}$ & $s$\\
\hline
\hline
match~$s_{i,j}$ in choice string~$c_{i,j}$ & $0$ & $k-2$ & $nk-2k+2$
& $nk-k$\\
\hline
template string~$t$ &\,$0$\,&\,$nk-k$\,&\,$0$\,&\,$nk-k$\,\\
\hline
\end{tabular}
\end{center}
As is obvious from these distance values, the
indicated substrings in the choice strings all have Hamming distance
$d=nk-k$ to the solution string and, therefore, are matches for~$s$.
\end{proof}

For the reverse direction, we assume that the $\CSS$ instance has a
solution. We need the following statements:

\begin{lem} \label{fronttag} A solution~$s$ and all its matches in the
input instance start with the front~tag.
\end{lem}

\begin{proof}
Since~$s$ is of length $L=(3nk+1)\cdot
nk+nk+{k\choose 2}(nk-2k+2)$, the only possible match in the template
string is the template string itself. Therefore, $s^{\prime}$ can
differ from $t^{\prime}$ in at most $d=nk-k$ symbols. We can show
that the only substrings in a choice string $c_{i,j}$ that
are possible matches for~$s$ with Hamming distance at most $d$ start with
the front tag, as we argue in the following.

Since~$s$ is a solution, there is a match in $c_{i,j}$ and we denote
it by~$s_{i,j}$. Denote the the first $(3nk+1)\cdot nk$ symbols of~$s_{i,j}$
by~$s^{\prime}_{i,j}$. Since $d_H(s^{\prime},s^{\prime}_{i,j})\leq
nk-k$ and $d_H(s^{\prime},t^{\prime})\leq nk-k$, we necessarily
(triangle inequality for Hamming metric) have
$d_H(s^{\prime}_{i,j},t^{\prime})\leq 2(nk-k)$. We show that this is
only possible when $s^{\prime}_{i,j}$ coincides with a front tag
of a block of~$c_{i,j}$. Assuming that it does not, we will show that
$d_H(s^{\prime}_{i,j},t^{\prime})> 2(nk-k)$, a contradiction. 

Firstly, assume that the starting position of $s^{\prime}_{i,j}$ and the
starting position of a front tag in~$c_{i,j}$ differ by $p$~positions,
$1\leq p\leq 3nk$. Then,
at least $nk-1$ symbols~``0'' of~$t^{\prime}$ are aligned with
symbols~``1'' of the front tag in~$s^{\prime}_{i,j}$ and at least $nk-1$
symbols~``1'' of~$t^{\prime}$ are aligned with symbols~``0''
of~$s^{\prime}_{i,j}$. This implies $d_H(s^{\prime}_{i,j},t^{\prime})>
2nk-2$. Secondly, assume that the starting position of
$s^{\prime}_{i,j}$ and the starting position of its closest front tag
in~$c_{i,j}$ differ by $p>3nk$~positions. Then,
a block of $3nk$ symbols~``1'' falls onto the encoding and/or the
back tag part of~$s^{\prime}_{i,j}$. Since the encoding part and back
tag contain together only $2+(nk-2k+2)<nk$ (under the assumption that
$k>2$) many symbols~``1'', we have more
than $2nk$ mismatching symbols and $d_H(s^{\prime}_{i,j},t^{\prime})>
2(nk-k)$.

Summarizing, we conclude that $s^{\prime}_{i,j}$ coincides with a
front tag in choice string~$c^{\prime}_{i,j}$, i.e.,
$s_{i,j}^{\prime}=t^{\prime}=s^{\prime}=\langle\mbox{front\_tag}\rangle$.
\end{proof}

\begin{lem} \label{exactlyk} The encoding part of~$s$ contains
exactly~$k$ symbols~``1''.
\end{lem}

\begin{proof}
Assume that~$s$ has less than~$k$
symbols~``1'' in its encoding part, i.e., $s^{\prime\prime}$ contains less
than~$k$ symbols~``1''. Then, because $t^{\prime\prime}=1^{nk}$,
$d_H(s^{\prime\prime},t^{\prime\prime})\geq nk-k+1$, implying
$d_H(s,t)\geq nk-k+1$, a contradiction.

Assume that~$s$ has more than~$k$ ``1''~symbols in its
encoding part~$s^{\prime\prime}$. Then,
$d_H(s^{\prime\prime},s_{i,j}^{\prime\prime})>k-2$ for the
encoding part~$s_{i,j}^{\prime\prime}$ of a match in every choice
string $c_{i,j}$. Now
consider the solution's back tag~$s^{\prime\prime\prime}$. To achieve
$d_H(s,s_{i,j})\leq nk-k$, we
need~$d_H(s^{\prime\prime\prime},s_{i,j}^{\prime\prime\prime})<nk-2k+2$
and $s^{\prime\prime\prime}$ must contain one or
more symbols~``1''. Every ``1''~symbol in~$s^{\prime\prime\prime}$
will decrease the value~$d_H(s,s_{i,j})$ for a block~$s_{i,j}$ of {\em
one} choice string $c_{i,j}$ by one, but will increase the
solution's Hamming distance to the selected blocks of {\em all other} choice
strings. No matter how many ``1''~symbols we have in the back tag of~$s$,
there will always be a choice string~$c_{i,j}$ with
$d_H(s^{\prime\prime\prime},s_{i,j}^{\prime\prime\prime})\geq
nk-2k+2$. In summary, we will always have a choice string~$c_{i,j}$ with
$d_H(s,s_{i,j})=d_H(s^{\prime\prime},s_{i,j}^{\prime\prime})+d_H(s^{\prime\prime\prime},s_{i,j}^{\prime\prime\prime})>nk-k$,
a contradiction.
\end{proof}

\begin{lem} \label{onepersec} Every section of the encoding part
of~$s$ contains exactly one symbol~``1''.
\end{lem}

\begin{proof}
Assume that not every section in the encoding
part of~$s$ contains exactly one ``1''~symbol. Then,
there must be a section containing no symbol~``1'', since, by
Lemma~$\ref{exactlyk}$, the number of symbols~``1'' in the
encoding part of~$s$ adds up to~$k$. Let $i^{\prime}$, $1\leq
i^{\prime}\leq k$, be the section containing no symbol~``1''.
W.l.o.g., consider a choice string~$c_{i^\prime,j}$, $i^{\prime}<j\leq
k$ or, if $i^{\prime}=k$, a choice string~$c_{i^\prime,j}$, $1\leq
j<i^{\prime}$. In {\em every} block~$s_{i^\prime,j}$ of~$c_{i^\prime,j}$,
sections~$i^{\prime}$ and~$j$ of the encoding part are active and, therefore,
contain exactly one symbol~``1'' each; these are the only symbols~``1''
in~$s_{i^\prime,j}''$. Now consider the $k$ symbols~``1'' in the
encoding part of~$s$: The ``1''s in all sections of~$s''$ except for
section~$j$ are all aligned with ``0''s in~$s_{i^\prime,j}''$; within
section~$j$,
only a single ``1'' of~$s''$ can be matched to a ``1''
of~$s_{i^\prime,j}''$. Therefore,
$d_H(s^{\prime\prime},s_{i^{\prime},j}^{\prime\prime})>k-2$. As in the
proof of Lemma~$\ref{exactlyk}$, we conclude that~$s$ is no solution.
\end{proof}


\begin{prop}
\label{reversek}
The~$k$ symbols~``1'' in the solution string's encoding part
correspond to a $k$-clique in the graph.
\end{prop}
\begin{proof} Let $s$ be a solution for the {\sc Closest Substring\/} instance.
Summarizing, we know by Lemma~$\ref{fronttag}$ that
$s$ can have as a match only one of the choice string's blocks.
By Lemma~$\ref{onepersec}$, every section of the
encoding part~$s^{\prime\prime}$ contains exactly one ``1'' symbol;
therefore, we can read this as an encoding of $k$~vertices of the
graph. Let $v_{h_1}, v_{h_2},\dots ,v_{h_k}$ be these vertices.
Further, we know that the back tag~$s^{\prime\prime\prime}$
consists only of ``0'' symbols: By Lemma~$\ref{exactlyk}$, the
encoding part~$s'$ has only $k$~``1''s; would $s^{\prime\prime\prime}$
contain a~``1'', then we would have $d_H(s,t)>nk-k$. We have
$d_H(s^{\prime\prime\prime},s_{i,j}^{\prime\prime\prime})= nk-2k+2$ for {\em
every} choice string match~$s_{i,j}$ and, since every~$s_{i,j}^{\prime\prime}$
contains only two ``1''~symbols,
$d_H(s^{\prime\prime},s_{i,j}^{\prime\prime})\geq k-2$. Now consider
some $1\leq i<j\leq k$ and the corresponding choice
string~$c_{i,j}$. Since $s$ is a solution, we know that there is a
block~$s_{i,j}$ with
$d_H(s^{\prime\prime},s_{i,j}^{\prime\prime})=k-2$.
That means that the two ``1''~symbols in $s_{i,j}^{\prime\prime}$ have
to match two ``1'' symbols in $s^{\prime\prime}$; this implies
that the two vertices $v_{h_i}$ and $v_{h_j}$ are
connected by an edge in the graph.
Since this is true for all $1\leq i<j\leq k$, vertices $v_{h_1},\dots ,
v_{h_k}$ are pairwisely interconnected by edges and form a $k$-clique.
\end{proof}

Propositions~$\ref{onedirk}$ and~$\ref{reversek}$ yield the following
main theorem:

\begin{thm}
$\CSS$ is $\W[1]$-hard for parameter~$k$ in the case of a binary alphabet.
\end{thm}

\section{{\sc Consensus Patterns\/}}
\label{mcpsec}

Our techniques for showing hardness of {\sc Closest
Substring\/}, parameterized by the number~$k$ of input strings, also apply
to {\sc Consensus Patterns\/}. Because of the similarity
to {\sc Closest Substring\/}, we restrict ourselves to explaining the problem and pointing out new features in the hardness proof.

Given strings $s_1, s_2,\dots , s_k$ over alphabet~$\Sigma$
and integers~$d$ and~$L$, the {\sc Consensus Patterns\/}
problem asks whether there is a string~$s$ of
length~$L$ such that~$\sum_{i=1}^{k}{d_H(s,s^{\prime}_i)}\leq d$
where $s^{\prime}_i$ is a length $L$ substring of $s_i$. Thus, {\sc
Consensus Patterns\/} aims for minimizing the {\em sum} of
errors. Since errors are summed up over all strings, the value of~$d$ will, usually, not be a small and, therefore, the most significant parameterization for this problem seems to be the one by~$k$. The problem is $\NP$-complete and has a PTAS~\cite{LMW99}. By reduction from {\sc Clique\/}, we can show
$\W[1]$-hardness results as for {\sc Closest Substring\/} given
unbounded alphabet size. We omit the
details here and focus on the case of binary input alphabet. We can apply
basically the same ideas as were used in Section~\ref{par_k}; however,
some modifications are necessary.

\subsection{Reduction of $\CLIQUE$ to {\sc Consensus Patterns\/}}
\label{cpconstr}
{\bf Choice strings.} As in Subsection~\ref{reduction_k}, we generate a
set of ${k\choose 2}$ {\it choice strings} $S_c=\{c_{1,2}$, $c_{1,2}
\dots, c_{k-1,k}\}$ with $c_{i,j}\ :=\
\langle\mbox{block}(i,j,e_1)\rangle
\langle\mbox{block}(i,j,e_2)\rangle
\dots
\langle\mbox{block}(i,j,e_m)\rangle$, encoding the~$m$ edges of the
input graph. This time, however, every block consists only of
a front tag and an encoding part. No back tag is necessary. Therefore,
we use $\langle\mbox{block}(i,j,(v_r,v_s))\rangle\ :=
\langle\mbox{front\_tag}\rangle\langle\mbox{encode}(i,j,(v_r,v_s))\rangle$,
in which the encoding part $\langle\mbox{encode}(i,j,(v_r,v_s))\rangle$ is
constructed as in Subsection~\ref{reduction_k}. Before we explain the
front tags, we already fix the distance value~$d$.

{\bf Distance Value.} We set the distance value $d:=({k\choose 2}-(k-1))nk$.

{\bf Front tags.} The front tag is now given by
$(1^{nk^3}0)^{nk^3}0^{nk^3}$. Thus, the front tag has
length~$n^2k^6+2nk^3$. The front tag here is more complex than the one used in
Subsection~\ref{reduction_k}. The reason is as follows. Its purpose is to make
sure that a substring which is not a block cannot be a match. To achieve this,
the front tag lets such an unwanted substring necessarily have a
distance value larger than~$d$ to a possible solution (as explained in
the proof of Lemma~$\ref{fronttag}$). Since~$d$ has a
higher value here compared to Section~\ref{par_k}, we need the more
complex front tag.

{\bf Solution length.} We set the substring length to the length of
one block, i.e., the sum of $n^2k^6+2nk^3$ (the length of the front
tag) and $nk$ (the length of the encoding part). Therefore,
$L:=n^2k^6+2nk^3+nk$.

{\bf Template strings.} In contrast to Subsection~\ref{reduction_k}, we
produce not only one but ${k\choose 2}-(k-1)$ many template strings. All
template strings have length~$L$, i.e., the length of one block. The template
strings are a concatenation of the front tag part (as given above) and an
encoding part consisting of $nk$ many symbols~``1''.

In summary, the front tag ensures that only the block of a choice
string can be selected as a substring matching a solution. Regarding
the distribution of mismatches, we note that a closest substring's front tag
part will not cause any mismatches. In its encoding part, every of its
$nk$~positions causes at least~${k\choose 2}-(k-1)$ mismatches. It
causes exactly~${k\choose 2}-(k-1)$ mismatches for every position
iff the input graph contains a $k$-clique. 

%

\subsection{Correctness of the Reduction}

\begin{prop}
\label{onedircp}
For a graph with a $k$-clique, the construction in Subsection~$\ref{cpconstr}$
produces an instance of {\sc Consensus Patterns} which has a solution, i.e.,
there is a string~$s$ of length~$L$ such that every $c_{i,j}$, $1\leq i<j\leq
k$, has a substring
$s_{i,j}$ with $\sum_{i=1}^{k-1}\sum_{j=i+1}^{k}d_H(s,s_{i,j})\leq d$.
\end{prop}

\begin{proof}
Given an undirected graph~$G$ with $n$~vertices and $m$~edges, let $1\leq
h_1<h_2<\dots<h_k\leq n$ be the indices of  $k$-clique's vertices. Then, let
string~$s$
consist of the front tag described in the above construction, concatenated
with the encoding part
$\langle\mbox{number}(h_1)\rangle\langle\mbox{number}(h_2)\rangle\dots
\langle\mbox{number}(h_k)\rangle$, which encodes all clique vertices. For
every $1\leq i<j\leq k$, we choose in
choice string~$c_{i,j}$ the block~$s_{i,j}$ encoding the edge connecting
vertices~$v_{h_i}$ and~$v_{h_j}$. We will show that these blocks have exactly
total Hamming distance $({k\choose 2}-(k-1))nk$ to~$s$.

The front tags of~$s$ and of each~$s_{i,j}$ coincide, their Hamming distance is
$0$. Recall from Subsection~\ref{reduction_k} that the encoding parts consist
of $k$ sections, each section of
length~$n$. We consider the encoding parts section by section and, within a
section, columnwise. Given a section~$i^{\prime}$, $1\leq
i^{\prime}\leq k$, there are $k-1$ choice strings in which this section is
active, and this section in these blocks encodes
vertex~$v_{h_{i^{\prime}}}$. Consider the column at position~$h_{i^{\prime}}$
in this section, over all selected substrings and all template strings. We
have ${k\choose 2}-(k-1)$ ``0''~symbols from the choice strings in which this
section is inactive; in all other strings, there is a ``1'' at this
position. In~$s$, this position is ``1,'' causing ${k\choose 2}-(k-1)$
mismatches. Now consider the remaining columns of section~$i^{\prime}$. In
each of them, we have
${k\choose 2}-(k-1)$ ``1''~symbols from the template strings; all
${k\choose 2}$ choice strings have~``0'' at the corresponding position. In~$s$,
this position is~``0,'' causing ${k\choose 2}-(k-1)$ mismatches. Thus,
we have ${k\choose 2}-(k-1)$ mismatches at
every of the $n$ positions within a section, and this is true for
all~$k$ sections of
the encoding part. The sum of distances from~$s$ to the matches in choice
strings and the template strings is~$({k\choose 2}-(k-1))kn$; ~$s$ is a
solution.
\end{proof}

For the reverse direction, we use two lemmas to show important properties
that a solution of the constructed instance has. The first lemma is proved in
analogy to Lemma~$\ref{fronttag}$.

\begin{lem} \label{fronttagcp} A solution~$s$ and all its matches in the
input instance start with the front~tag.\hfill\qed
\end{lem}

The second property of a solution, although also valid for the solutions in
Subsection~\ref{correctnessk}, is established in a different way here. It
relies on the additional template strings that have been introduced in the
construction of the {\sc Consensus Patterns\/} instance.

\begin{lem}
\label{onepositions}
A solution~$s$ contains exactly one symbol~``1'' in every section of its
encoding part.
\end{lem}

\begin{proof}
Let $s$ be a solution for the constructed {\sc Consensus Patterns\/} instance.
By Lemma~$\ref{fronttagcp}$, we know that~$s$ and all
its matches in the choice strings start with the front
tag. Consequently, the matches in the choice strings must be blocks.

Consider the encoding part of a solution~$s$ together with the encoding parts
of its matches in the input strings. We note that we
have at least ${k\choose 2}-(k-1)$ mismatches for every column at
positions~$p$, $1\leq p\leq nk$: On the
one hand, all ${k\choose 2}-(k-1)$ template strings have ``1''~symbols
at position~$p$. On the other hand, all ${k\choose 2}-(k-1)$~choice
strings in which position~$p$'s section is inactive have ``0''
at this position, no matter which blocks we chose in these choice
strings. Since $s$~is a solution and only a total of~$({k\choose 2}-(k-1))nk$
mismatches are allowed, we have {\em exactly} ${k\choose 2}-(k-1)$
mismatches for every position of the encoding part of~$s$ with the
corresponding positions in the matches of~$s$.

Now, consider an arbitrary section~$i^{\prime}$, $1\leq i^{\prime}\leq
k$, and consider all $k-1$ choice strings in which
section~$i^{\prime}$ is active. In these choice strings, section~$i^{\prime}$
contains exactly one ``1''~symbol. We will show that in these
choice strings' blocks that form the matches for~$s$, the ``1'' in
section~$i^{\prime}$ must be at the same position in all matches, because,
otherwise, $s$ is no solution. Assume that we chose blocks in which
the ``1''~symbols of section~$i^{\prime}$ are at different positions. We can
easily check that this would cause more than ${k\choose 2}-(k-1)$ mismatches
for the columns corresponding to the positions of the ``1''~symbols; this
contradicts the assumption that~$s$ is a solution. We conclude that, for all
matches in choice strings, the ``1''~symbols of section~$i^{\prime}$ must be
at the same position. For columns in which we have ``1''~symbols in choice
strings, there is a majority of ``1'' symbols, namely those in the $(k-1)$
choice strings in which section~$i^{\prime}$ is active and those in the
${k\choose 2}-(k-1)$ template strings. Therefore, the
respective position in~$s$ must be ``1.'' For all other columns, there is a
majority of ``0'' symbols, namely those in all ${k\choose 2}$ choice
strings. Therefore, the respective position in~$s$ must be~``0.''
\end{proof}

These two lemmas allow us to show that also the reverse direction of the
reduction is correct. 

\begin{prop}
\label{reversecp}
The~$k$ symbols~``1'' in the solution string's encoding part
correspond to a $k$-clique in the graph.
\end{prop}

\begin{proof}
Let $s$ be a solution for the constructed {\sc Consensus Patterns\/} instance.
By Lemma~$\ref{onepositions}$, every section in the encoding part of~$s$
encodes a vertex of the input graph. In the following, we show that all
encoded vertices are interconnected by edges.

Let $V_C=\{v_{h_1}, v_{h_2},\dots ,v_{h_k}\}$ be
the vertices encoded in the solution's encoding part. For every two
sections $1\leq i<j\leq k$, we select in choice string~$c_{i,j}$ a substring
in which the ``1'' symbols of sections~$i$ and~$j$ are at the same positions
as the ``1'' symbols of sections~$i$ and~$j$ in the solution: Selecting
another substring would result in a Hamming distance greater than ${k\choose
  2}-(k-1)$ in the $h_i$th and $h_j$th column and~$s$ could not be a solution.
Hence, the selected block encodes the edge
connecting~$v_{h_i}$ and~$v_{h_j}$. Since we find such a substring for
every $1\leq i<j\leq k$, every pair of vertices in~$V_C$ is connected
by an edge, $V_C$ is a $k$-clique.
\end{proof}

Propositions~$\ref{onedircp}$ and~$\ref{reversecp}$ yield the following main
result. 

\begin{thm}
\label{mcpthm}
{\sc Consensus Patterns\/} is $\W[1]$-hard for parameter~$k$ in case
of a binary alphabet.
\end{thm}   

\section{Conclusion}\label{concl}

We have proven that {\sc Closest Substring\/} and {\sc Consensus Patterns},
parameterized by the number~$k$
of input strings and with alphabet size two, are $\W[1]$-hard.
This contrasts with related sequence analysis problems, such as
{\sc Longest Common Subsequence}~\cite{BDFW95,BDFHW95}
and {\sc Shortest Common Supersequence}~\cite{Hal96}, where, until now,
parameterized hardness has only been established in the case
of unbounded alphabet size.
Now, it is also known that these problems, parameterized by the number of
input strings, are W[1]-hard in case of bounded alphabet
size~\cite{Pie02}.
In our opinion, however, intuitively speaking, our W[1]-hardness 
result for {\sc Consensus Patterns} is the most surprising one in this
context, because {\sc Consensus Patterns} seems to carry significantly
less combinatorial structure than the other problems.

The parameterized complexity of  {\sc
Closest Substring\/} and {\sc Consensus Patterns\/}, parameterized by
``distance parameter''~$d$,
remains open for alphabets of constant size.  If these problems are
also $\W[1]$-hard, then an efficient and practically useful
PTAS would appear to be impossible \cite{CT97,DF99},
unless further structure of natural input distributions is taken into account in
a more complex aggregate parameterization of these basic computational problems of bioinformatics.

\bibliographystyle{plain}

\end{document}